\documentclass[aps,pre,superscriptaddress,a4paper,floatfix,nofootinbib,twocolumn]{revtex4-1}
\pdfoutput=1
\usepackage{graphicx,color,bigints,amsmath,amssymb,enumerate}%amssymb,
\usepackage[normalem]{ulem}
\usepackage[urw-garamond]{mathdesign}
\usepackage[T1]{fontenc}
\usepackage[ocgcolorlinks,colorlinks=true,linkcolor=blue,citecolor=red]{hyperref}

\newcommand{\tr}{\textrm{tr}}
\newcommand{\bra}[1]{\langle #1|}
\newcommand{\ket}[1]{|#1\rangle}
\newcommand{\Exp}[1]{\langle#1\rangle}
\newcommand{\braket}[2]{\langle #1 | #2 \rangle}
\newcommand{\kets}[2]{|#1\rangle_{\!_#2}}
\newcommand{\bras}[2]{{}_{_#2\!\!}\langle#1|}
\newcommand{\proj}[2]{|#1\rangle_{\!_#2\!\!}\langle#1|}

\newcommand{\delp}{\delta \!\!p}
\newcommand{\delW}{\delta \!W}
\newcommand{\delF}{\delta \!F}

\begin{document}
\title{Extracting work from quantum systems}
\author{Paul Skrzypczyk}\affiliation{Department of Applied Mathematics and Theoretical Physics$\text{,}$ University of Cambridge, Centre for Mathematical Sciences, Wilberforce Road, Cambridge CB3 0WA, United Kingdom}
\author{Anthony J.~Short}\affiliation{H. H. Wills Physics Laboratory, University of Bristol$\text{,}$ Tyndall Avenue, Bristol, BS8 1TL, United Kingdom}
\author{Sandu Popescu} \affiliation{H. H. Wills Physics Laboratory, University of Bristol$\text{,}$ Tyndall Avenue, Bristol, BS8 1TL, United Kingdom}

\begin{abstract}
	We consider the task of extracting work from quantum systems in the resource theory perspective of thermodynamics, where free states are arbitrary thermal states, and allowed operations are energy conserving unitary transformations. Taking as our work storage system a `weight' we prove the second law and then present simple protocols which extract \emph{average} work equal to the free energy change of the system -- the same amount as in classical thermodynamics. Crucially, for systems in `classical' states (mixtures of energy eigenstates) our protocol works on a single copy of the system. This is in sharp contrast to previous results, which showed that in case of \emph{almost-deterministic} work extraction, collective actions on multiple copies are necessary to extract the free energy. This establishes the fact that free energy is a meaningful notion even for individual systems in classical states.  However, for non-classical states, where coherences between energy levels exist, we prove that collective actions are necessary, so long as no external sources of coherence are used.
\end{abstract}

\maketitle

\section{Introduction}
Thermodynamics forms part of the bedrock of our current understanding of the physical world. It has remained unchanged despite huge revolutions in physics, such as relativity and quantum theory, and few believe it will ever fail. Over time, it has been applied to situations well outside its original domain; from black holes \cite{Bek73,BarCarHaw73}, to quantum engines comprised of only a few qubits \cite{GevKos96,YouMahOba09,TonMah05,LinPopSkr10a}. A fundamental question is that of the applicability of thermodynamics to quantum systems \cite{GemMicMah04,GyfBer05}; it is this question that we wish to address in this paper.
 
One of the most fundamental questions one would like to ask is how much work can be extracted from a system, given access to a thermal bath at temperature $T$. This question, for generic quantum systems and Hamiltonians has been investigated intensively, e.g. in \cite{ProLev76,AliHorHor04,HasIshDri10,TakHasDri10,EspVan11,Abe11,VerLac12} yet, until very recently only partial results have been obtained and it remained an open problem. Very recently however strong general results were obtained \cite{HorOpp11}, made possible by employing a new conceptual framework, namely that of ``resource theories".

The advent of resource theories is one of the main paradigm shifts which occurred within quantum information theory; the main idea is to bring to the forefront the study of \emph{resources} needed to implement various tasks. The archetypal resource theory in this context is entanglement theory \cite{BenBerPop96,HorHorHor09}. However, numerous other examples have now been developed, including reference frames \cite{BarRudSpe07}, purity \cite{HorHorOpp03} and asymmetry \cite{MarSpe11}, which has culminated in a general formalised notion of a resource theory \cite{HorOpp12}. In \cite{BraHorOpp11} the authors introduced an appealing resource theory of thermodynamics, based on the class of allowed operations first studied in \cite{JanWocZei00}.

Here we revisit and modify the resource theory framework of thermodynamics as originally put forward in \cite{BraHorOpp11}.  In this new context we achieve two main results: First we prove the second law of thermodynamics.  Second, as described in more detail below, we show that free energy is a relevant quantity for individual quantum systems. 

The latter result is particularly relevant following the results of \cite{HorOpp11} which call into question the role of free energy for individual quantum systems. Classical thermodynamics tells us that the total amount of work we are able to extract from a system is given by its change in \emph{free energy}, which was also supported by previous partial quantum results \cite{ProLev76,AliHorHor04,HasIshDri10,TakHasDri10,EspVan11,Abe11,VerLac12}. Yet in \cite{HorOpp11} it was shown that work equal to free energy can be extracted only if we {\it collectively} process many copies of the same system. When acting on each copy individually, the amount of work that can be extracted is generally significantly less than the free energy. These results therefore suggest that the free energy is not the relevant quantity for individual systems.

Here we revisit the issue of work extraction and show that free energy is a significant quantity for individual systems. Our paradigm is similar to that of \cite{HorOpp11} but differs in an essential aspect. In \cite{HorOpp11} they considered {\it almost deterministic}\footnote{Specifically, they require that a deterministic amount of work is extracted, except with some small probability $\epsilon$ of failure} work extraction, from the `single-shot' viewpoint which has received much attention lately \cite{DahRenRie11,Abe11,RioAbeRen11,EglDahRen12,FaiDupOpp12}. Here, in contrast, we will consider {\it average} work extraction. We prove that there exist simple protocols which extract \emph{average} work equal to the free energy change of the system. Crucially, for systems in `classical' states (mixtures of energy eigenstates) our protocol works on a single copy of the system. This establishes the fact that free energy is a meaningful notion even for individual systems, when their states are classical.

However, for non-classical states, where coherences between energy levels exist, we prove that collective actions are necessary, so long as no external sources of coherence are used. More precisely, we prove that when acting on an individual system the amount of average work that can be extracted is equal to the free energy of its decohered state -- the free energy due to coherences is `locked' and cannot be used. As more copies are collectively processed, some of this locked free energy is released -- we can extract the free energy of the decohered state of $n$ copies, which is larger than $n$ times the free energy of a single decohered system.  In the limit of infinitely many copies, the average work extracted per copy reaches its entire free energy.  Whether or not external coherent resources allow extraction of work equal to the free energy for individual systems, and how to account for such resources (given that they may provide free energy themselves even without the system) remains an open question.

The paper is organised as follows. In Sec.~\ref{s:preliminaries} we go through the preliminaries, including a brief description of the resource theory of thermodynamics in Sec.~\ref{ss:resource theory}, work storage systems in Sec.~\ref{ss:work storage} and a proof that the second law holds in Sec.~\ref{s:second_law}. In Sec.~\ref{s:diagonal} we study work extraction from diagonal systems, beginning with a motivating example in Sec.~\ref{s:motivating example}, before moving onto a full qubit protocol in Sec.~\ref{ss:full protocol}. We discuss a special case in Sec.~\ref{ss:maxwell demon}, present a full protocol for arbitrary dimension systems in Sec.~\ref{ss:qudits} and prove optimality in Sec.~\ref{s:optimality}. In Sec.~\ref{s:coherence} we show how work can be extracted from coherence by acting collectively on multiple copies of the system. We finish by briefly discussing how the results of this paper can be understood from the perspective of `virtual temperatures' in Sec.~\ref{s:virtual}, before concluding in Sec.~\ref{s:conclusions}.

\section{Preliminaries}\label{s:preliminaries}
\subsection{The resource theory of thermodynamics}\label{ss:resource theory}
The first ingredient in a resource theory is a set of free resources. By free we mean that we have an unlimited supply of this resource, since it is deemed easy to acquire. In the present context of thermodynamics this free resource is \emph{a thermal bath at temperature $T$}. More precisely, we assume that we have an unlimited supply of finite-dimensional systems, with any desired Hamiltonian $H$,  in their thermal state $\tau_\beta(H)$,
\begin{equation}
	\tau_\beta(H) = \tfrac{1}{\mathcal{Z}}e^{-\beta H}
\end{equation}
where $\beta = 1/T$ is the inverse temperature\footnote{We set $k_B=1$ throughout for convenience}, and $\mathcal{Z}~=~\tr(e^{-\beta H})$ is the partition function.

The second ingredient is a set of allowed operations, those which one has the ability to perform with ease. In the original resource framework these allowed operations were \emph{any energy conserving unitary transformation}. In particular, any unitary operation which commutes with the total (non-interacting) Hamiltonian of all the systems under consideration. This allows one to transform systems in a controlled way, but without admitting an external source of energy.

However, we will also introduce an additional element into  the resource framework, a \emph{work storage system}, whose sole purpose is to provide and store work for thermodynamic transformations. In the next section, we will place some additional constraints on the allowed operations, which prevent us from `cheating' by  using work storage system for other purposes (e.g. by using it  as a cold reservoir in a heat engine), and allow us to use the same work storage system for multiple thermodynamic protocols.

Given these ingredients, any additional system which is not in thermal equilibrium with a bath at temperature $T$ is a useful resource. The question that remains is to quantify how useful different resource states are.

\subsection{Work storage systems}\label{ss:work storage}
One way to quantify the usefulness of different resource states is to find a transformation from every resource state to some \emph{fiducial} resource which allows for comparison in a well-defined and easy manner. In the resource theory of entanglement, for example, one can quantify the entanglement of a state by the optimal asymptotic number of maximally entangled states that can be created per resource state, or in other words the optimal rate at which states can be converted into maximally entangled states. Different states will have different optimal rates, and thus the rate quantifies how useful they are. We seek here an  analogous fiducial resource for thermodynamics.

Work is the desired output in this case, and different states can produce work at different rates. Because we wish to study a closed quantum system, with  conserved total energy, we desire a system where we can store the work extracted from resource states. Similarly to entanglement theory, where the maximally entangled state is chosen for convenience, here we wish to find a work storage system which we deem to be most convenient.

In previous work \cite{HorOpp11} a candidate was put forward -- raising a qubit deterministically from its ground state to it's excited state. This qubit was termed a \emph{wit}, short for \emph{work bit}. However, choosing the energy gap of the work bit requires advance knowledge of the work to be extracted, and this model does not translate well to non-deterministic work extraction, which we will be interested in here. Furthermore, we would prefer to be able to use a single work storage system as a `battery' capable of gaining and expending work in multiple thermodynamic processes.

The alternative work storage system we will consider here is a suspended \emph{weight}, which is raised or lowered when work is done on or by it. In particular, we will consider a quantum system whose height is given by the position operator $\hat{x}$, with Hamiltonian $H_w= m g \hat{x}$ representing its gravitational potential energy\footnote{Note that it is not strictly necessary for the height of the weight to be a continuous degree of freedom. In Appendix \ref{a:discrete work} we show that the weight can be replaced by a system with discrete equally-spaced energy levels. This introduces a small error in the results, but this can be made as small as desired by appropriately scaling the spacing of the weight.}.  For simplicity, we choose $mg =1 \textrm{ J/m}$, such that the value of  $\hat{x}$ directly denotes the work stored by the mass.  Such a system has a long history of being used as a work storage system in classical thermodynamics \cite{LieYng99}.

We will place two additional constraints on the allowed dynamics governing interactions with the weight:\renewcommand{\labelenumi}{(\roman{enumi})}
\begin{enumerate}
\item The average amount of work extracted in an allowed protocol must be independent of the initial state of the weight.\label{i:I}

\item All allowed unitaries should commute with translation operations on the weight. This reflects the translational symmetry of the weight system, and the fact that only displacements in its height are important.\label{i:II}
\end{enumerate}

These constraints are intended to prevent us from `cheating' by using the weight for purposes other than as a work storage system (e.g. as a cold reservoir, or a source of coherence). For example if the weight starts in a pure state, it has zero entropy, so one needs to be careful to avoid unwittingly using this as a source of free energy, in addition to the system. Constraint (i) encapsulates this requirement. 

Constraint (ii) is a bit more subtle. On the one hand, it is very plausible that it is a consequence of constraint (i). We however have not been able to show this. On the other hand, even if it does not follow from (i), we would still like to assume it, as it seems wrong to use properties of the weight in the protocol, even if they do not affect the work extracted.

Furthermore, constraint (i) ensures that we can use the same work storage system for multiple thermodynamics protocols (or on several copies of the same state) without having to worry how the initial state has been modified by earlier procedures. 

As we are free to choose any initial state of the weight due to constraint (i) above, for simplicity we will usually take the initial state of the weight to be a narrow normalised wavepacket $\ket{0}_w$ centred on the origin (i.e. $\bra{0} \hat{x} \ket{0} =0$). When the same wavepacket is raised above the origin by a distance $a$, we will denote the state by $\ket{a}_w=\Gamma_a \ket{0}_w$, where $\Gamma_a$ is the translation operator\footnote{The translation operator is $\Gamma_a=\exp(-i a \hat{p} / \hbar)$, where $\hat{p}$ is the usual momentum operator satisfying  $[\hat{x}, \hat{p}] = i \hbar$. }.

\section{The second law} \label{s:second_law}
We now show that the second law of thermodynamics holds in our  framework, by proving a probabilistic version of the second law -- that there is no protocol which extracts a positive quantity of average work from a thermal bath (i.e. that there is no way on average of turning heat into work) \cite{LadPreSho08}. To show this we will use proof by contradiction. 

Consider a thermal bath at temperature $T$ and the weight, i.e. there is no additional system out of thermal equilibrium. 

Let us first consider the energy changes during the protocol. Suppose that we are able to extract work from the bath and store it in the weight, $\Delta E_W > 0$. The average energy of the thermal bath must change by $\Delta E_B = - \Delta E_W $ due to energy conservation.

Now consider the entropy changes during the same protocol (in particular, the changes in von Neumann entropy $S(\rho) = - \tr (\rho \log \rho))$. As the bath and weight are initially uncorrelated, their initial entropy is simply the sum of their individual entropies.  Unitary transformations conserve the total entropy, $\Delta S_{BW} = 0$. However, as correlations can form during the protocol, the sum of the final entropies of the bath and weight can be greater than the sum of their initial entropies (as the entropy is subadditive). This means that
\begin{equation}\label{e:entropy eqn}
	\Delta S_B + \Delta S_W \geq \Delta S_{BW} = 0
\end{equation}

Given an initial thermal state for the bath (with positive temperature), any change of the state which reduces its average energy must also reduce its entropy\footnote{This follows from the fact that the thermal state is the maximal entropy state with given average energy. For more details see Appendix \ref{a:second_law}}, $\Delta S_B < 0$. However, within our framework all allowed protocols are such that the average work extracted is independent of the initial state of the weight; we are therefore free to choose any initial state of the weight we like. Consider now that the initial state of the weight is a very wide wavepacket. We show in Appendix \ref{a:second_law} that the entropy change of the weight in this case can be made as small as desired, in particular we can make $\Delta S_W < |\Delta S_B|$. This would result in violating \eqref{e:entropy eqn}. Hence there is a contradiction, and thus there is no way to extract work from the bath.

\section{Extracting work from diagonal systems}\label{s:diagonal}
We will begin our exploration of work extraction by considering the special case of extracting work from systems whose states are diagonal in the energy eigenbasis. This represents systems which are essentially classical, since coherence plays no role, and states are simply mixtures of energy eigenstates. We will present a protocol which extracts work equal to the change in free energy of the system. In contrast to existing protocols, here our protocol will take one copy of the system, and extract average work equal to the free energy of the system.

The protocol contains at its core a single elementary building block which is able to extract a vanishing amount of work blue, with maximal efficiency. The full protocol consists of many repetitions of this one step. We shall present first this building block in a motivating example and then show precisely how it is used to build the full protocol. In the main text the emphasis will be on the results of the protocol, with the more detailed analysis provided in the appendices.

\subsection{A motivating example}\label{s:motivating example}
Let us consider the task of performing the following transformation on a single qubit
\begin{eqnarray}\label{e:rhoS}
	\rho_S &=& (1-p)\proj{0}{S} + p \proj{1}{S} \nonumber \\
	&\downarrow \\
	\rho_S' &=& (1-p+\delp)\proj{0}{S} + (p-\delp)\proj{1}{S} \nonumber
\end{eqnarray}
whose Hamiltonian is $H_S = E_S \proj{1}{S}$ and where $\delp \ll 1$. That is, we wish to consider the small transformation from a diagonal state to a nearby diagonal state. The initial and final state differ in free energy,  and hence the goal is to extract an amount of work as close as possible to this difference, namely
\begin{align}
	\delF_\beta &= F_\beta(\rho) - F_\beta(\rho') \nonumber \\
	&= \left(\Exp{E}_{\rho} - T S(\rho)\right) - \left(\Exp{E}_{\rho'} - T S(\rho')\right) \nonumber \\
	&\simeq \delp \left(E_S - T S_c'(p)\right)
\end{align}
where $F_\beta(\sigma)=\Exp{E}_{\sigma} - TS(\sigma)$ is the free energy of $\sigma$ with respect to a bath at inverse temperature $\beta$, $\Exp{E}_\sigma = \tr\left(H_S \sigma\right)$ is the average energy, $S_c(q)~=~-q \log q~-~(1-q)\log(1-q)$ is the (classical) binary entropy, $S_c'(q) = \frac{dS_c}{dq}$, and we have taken the first order expansion in $\delp$ to arrive at the final line. We shall now see that to first order we can extract exactly this amount of work from the system.

To do so let us bring in a thermal state from the bath -- the free resource at our disposal -- and choose its Hamiltonian such that the population in the excited state is $p-\delp$. For thermal qubits the relation between excited state population $r$ and energy spacing $E_B$ is
\begin{equation}
	E_B = T\log\left(\frac{1-r}{r}\right) = TS_c'(r)
\end{equation}
thus we will choose the Hamiltonian of the bath qubit to be $H_B = E_B\proj{1}{B}$ with
\begin{align}
	E_B &= T S_c'(p-\delp) \simeq T S_c'(p) - \delp T S_c''(p)
\end{align}
Let us take the weight to be prepared initially in the state $\kets{0}{w}$ of zero average energy, and consider applying an energy conserving unitary transformation $U$ which interchanges
\begin{equation}\label{e:U mot}
	\kets{0}{B}\kets{1}{S}\kets{x}{w} \leftrightarrow \kets{1}{B}\kets{0}{S}\kets{x+E_S-E_B}{w}
\end{equation}
for any $x$, whilst leaving all orthogonal states unchanged\footnote{$U$ can be expressed in terms of the translation operator $\Gamma_a$ as 
\begin{equation*}
U=\ket{10}\bra{01} \otimes\Gamma_{\Delta } + \ket{01}\bra{10} \otimes\Gamma_{-\Delta } + \left(\ket{00}\bra{00} + \ket{11}\bra{11}\right)\otimes\openone,
\end{equation*} where $\Delta = E_S -E_B$. Note that this commutes with the free Hamiltonian $H_S+H_B+H_w$.}.

It is the change in average energy of the weight which is of interest. A straightforward calculation (given for completeness in the Appendix) shows that this is given by
\begin{align}
	\delta \Exp{E}_w = \delW &= \delp(E_S - E_B) \simeq \delp(E_S - T S_c'(p))
\end{align}
which is equal to $\delF_\beta$ to first order in $\delp$.

Thus we observe that if we wish to to make a change of state from $\rho$ to $\rho'$, a way to proceed is to bring in a qubit from the bath in the state $\tau_\beta(H_B) = \rho'$. Whenever $\rho$ is not the thermal state (given it's Hamiltonian $H_S$), we find that $H_B \neq H_S$, and therefore there is a difference in energy between the excited states of the two qubits. It is thus not possible to swap the state of the two qubits using an energy conserving unitary. However, by bringing in the weight we can perform the unitary \eqref{e:U mot} which is energy conserving, and, at the level of the reduced states of the system and bath, indeed performs a swap. The system therefore ends up in the desired state $\rho'$, and we find that the change in average energy of the weight, when the state of the system and bath are close, approaches exactly the change in free energy of the system.

\subsection{Full qubit protocol}\label{ss:full protocol}
The above insight can be used to build a full protocol which is able to extract an amount of work approaching the free energy change in transforming a system from a state $\rho$ to its thermal state $\tau_\beta(H_S)$.

The basic idea is that, since in making a small change of state we have seen that the work extracted approaches the free energy change, we can divide a large change of state into many small changes and at each stage extract the desired amount of work.

Therefore, let us fix an integer $N \gg 1$, and consider $N$ thermal qubits $\tau_\beta^{(k)}(H_B^{(k)})$, $k = 1,\ldots,N$,
\begin{equation}
	\begin{split}
		\tau_\beta^{(k)}(H_B^{(k)}) &= (1-r^{(k)})\proj{0}{{B_k}} + r^{(k)}\proj{1}{{B_k}}\\
		r^{(k)} &= p + \tfrac{k}{N}(p_{eq}-p)
	\end{split}
\end{equation}
where $p_{eq} = \frac{e^{-\beta E_s}}{1+e^{-\beta E_s}}$ is the excited-state probability in the thermal state $\tau_\beta(H_S)$. This collection of bath qubits have excited-state probabilities which vary linearly from that of the initial state of the system $\rho_S$ to that of the thermal state of the system $\tau_\beta(H_S)$, in increments of $\tfrac{(p_{eq}-p)}{N}$. We will apply, in turn, the sequence of unitary transformations $\{U^{(k)}\}$, where $U^{(k)}$ acts on bath qubit $k$, system and weight and swaps the states
\begin{equation}
	\kets{0}{{B_k}}\kets{1}{S}\kets{x}{w} \leftrightarrow \kets{1}{{B_k}}\kets{0}{S}\kets{x+E_S-E_B^{(k)}}{w}
\end{equation}
whilst leaving all other orthogonal states unchanged.

Assuming once again that the weight starts in the state $\kets{0}{w}$, it is the final state, and in particular the final average energy of the weight which is of interest, as the number of steps $N$ tends to infinity. In Appendix \ref{a:weight} we calculate explicitly this state in the asymptotic limit. The result is that the final reduced state in the infinite limit, denoted $\Omega_w^\infty$, is
\begin{multline}\label{e:asymptotic ladder}
	\Omega_w^\infty =
	(1-p)\proj{T\log\left(\tfrac{1-p}{1-p_{eq}}\right)}{w} \\+ p\proj{T\log\left(\tfrac{p}{p_{eq}}\right)}{w}
\end{multline}
The average energy is thus
\begin{align}
	\Exp{E}_w^\infty &= T\left(p \log\left(\tfrac{p}{p_{eq}}\right) + (1-p) \log\left(\tfrac{1-p}{1-p_{eq}}\right)\right) \nonumber \\
	&= T D_c(p||p_{eq}) = F_\beta(\rho) - F_\beta\left(\tau_\beta(H_S)\right) 
\end{align}
where $D_c(p||p_{eq})$ is the relative binary entropy.

We observe that the average energy of the weight at the end of the protocol, and hence the average work extracted,  approaches precisely the free energy difference in the asymptotic limit of infinitely many thermal qubits. Although this is promising, it should be noted that the final distribution of the weight does not have the form one may intuitively have asked for (i.e. a peak around a single value, at the free energy itself). Instead we find that the solution has perfect correlation between the initial state of the system and the final energy of the weight. It is worth noting that this interesting distribution coincides with the results found in \cite{Abe11} where, even though the details of the allowed transformations differ, it was shown that for classical systems, optimal work extraction protocols converge on this distribution in probability.

Note also that in the large $N$ limit, although many thermal qubits are required, each of them is only slightly perturbed by the process.

Finally, we consider the case of $n$ independent copies of the system -- i.e. the macroscopic limit. To do so we simply repeat the protocol above on each qubit separately, using a single weight to store the work. In each case we have seen that the weight is transformed from a single peak (at $\kets{0}{w}$ above) to two peaks. Since the protocol is linear, it follows immediately the the final distribution of the weight will be the $n$-fold convolution of the above distribution for a single qubit. From the central limit theorem one concludes that the final average energy of the weight is
\begin{equation}
	\Exp{E}_w = n \left(F_\beta(\rho) - F_\beta(\tau_\beta(H_S))\right)
\end{equation}
with fluctuation of the order $\sqrt{n}$. For large $n$ these fluctuations will be insignificant compared to the average, and we obtain an almost deterministic quantity of work, with the amount of work extracted per copy of the system equal to the free energy difference.

The results presented here are completely consistent with those presented in \cite{BraHorOpp11}. It should be noted however the stark difference in which the two protocols achieve the same goal. In \cite{BraHorOpp11} the analysis first considers all $n$ systems together and uses ideas from information theory to compress information within large typical subspaces. Here, on the other hand, each qubit is manipulated separately and nowhere is the idea of a typical subspace ever introduced, nevertheless they both achieve the same end result.

\subsection{Special case: isothermal expansion}\label{ss:maxwell demon}
The protocol outlined in the previous subsection is the general protocol which allows for the extraction of work from an arbitrary diagonal state of a qubit, for any given Hamiltonian. There is one particular choice of state that will be particularly important in later sections. As such, we shall briefly look at it in more detail.

Consider a qubit for which the Hamiltonian vanishes ($E_S = 0$) and for which all the probability is initially in the state $\kets{0}{S}$ ($p = 0$). Since the Hamiltonian vanishes, the system lives in a degenerate space, and since it is pure we have maximal information about the state of the system. This information can be used to extract work from the heat bath. Since all states of the system have equal energy, it is clear that the system provides only information, whilst it is the bath which provides the energy.

For a degenerate system the equilibrium state $\tau_\beta(0) = \tfrac{1}{2}\openone$, thus after applying the protocol from the previous section, in the asymptotic limit of infinitely many bath qubits, the final state of the weight is given by
\begin{equation}
		\Omega_w^\infty = \proj{T\log 2}{w}
\end{equation}
corresponding to the extraction of an amount of work $W = T\log 2$ from the system. This is an analogous process to the isothermal expansion of a single-molecule gas from one side of a box into the full volume, as discussed in Szilard's well-known treatment of Maxwell's demon \cite{Szi29,LefRex03}.

It should be noted that contrary to the previous subsection, where the final state of the weight was found to consist of two peaks, in this special case of a pure initial state it consists of only a single peak. Thus it is not necessary to go to the many-copy regime to extract a deterministic amount of work in this case.

\subsection{Extracting work from diagonal quantum states of arbitrary dimension }\label{ss:qudits}

To conclude the analysis of diagonal systems it is now necessary to move beyond qubits, to quantum systems of arbitrary finite dimension $d$, and arbitrary Hamiltonians.

Our strategy is to connect a pair of energy levels in the system to a sequence of thermal qubits in the bath, shifting the occupation probability between these levels as in the qubit case (Sec.~\ref{ss:full protocol}). We then repeat this process for different pairs of levels until we have the desired final state. As the amount of work extracted in each step does not depend on the initial state of the weight (as proven in Appendix \ref{a:work_independence}), we can consider each step separately and simply sum the total work extracted. 

An important difference from the qubit case is that the occupation probability of the two levels in the system that are coupled to the bath at any given stage need not sum to unity. For example consider a qutrit (or larger system) in the state
\begin{equation}
	\rho_S = (1-p-q)\proj{0}{S}+q\proj{1}{S} + p\proj{2}{S},
\end{equation}
where we wish to shift a small amount $\delp$ of probability from $\kets{0}{S}$ to $\kets{2}{S}$.  To do this, we take a thermal qubit with excitation probability $(p+\delp)/(1-q)$,  so that the \emph{ratio} of occupation probabilities in the bath qubit is the same as in that in the desired final state of the system. We then apply the unitary
\begin{equation}\label{e:U mot2}
	\kets{0}{B}\kets{2}{S}\kets{x}{w} \leftrightarrow \kets{1}{B}\kets{0}{S}\kets{x+E_2-E_B}{w}
\end{equation}
where $E_2$ is the energy of the state $\kets{2}{S}$, and all orthogonal states are left unchanged. We show in Appendix \ref{a:qudits} that the change in the average energy of the weight during this process is equal to the free energy change of the system, to first order in $\delp$.

From this it is straightforward to extend the qubit protocol to arbitrary diagonal states. For sufficiently large $N$ we can choose a sequence of  $N+1$ states for the system where two adjacent states differ in occupation probability between a pair of levels by $\mathcal{O}(1/N)$, with the first and last states equal to the initial state of the system and its thermal state respectively\footnote{For example, we could first shift probability from all energy levels with higher probability in $\rho_S$ than in $\tau_\beta(H_S)$ to the $\kets{0}{S}$ state, then move probability from  $\kets{0}{S}$  to the remaining levels, using $N/(d-1)$ steps for each pair of levels.}. We then pick the Hamiltonian for $N$  thermal qubits such that unitaries of the form \eqref{e:U mot2} take the reduced state of the system from from one state in the sequence to the next. This is done by taking the thermal qubits to have populations equal to those of the pair of levels of interest of the next state in the sequence, after renormalising. In the limit  $N \rightarrow \infty$ the average amount of work put into the weight equals the free energy change of the system, regardless of the precise choice of path.

Interestingly, an equivalent protocol would work for transitions between any two diagonal states, $\rho$ and $\sigma$, extracting an average amount of work equal to $F_\beta(\rho) - F_\beta(\sigma)$. This implies that creating a diagonal state $\rho$ from a thermal state costs the same average amount of work as can be extracted from $\rho$ when it is thermalised. In this sense thermodynamic transitions between diagonal  states are reversible \footnote{However, note that if a state is thermalised and then recreated using our protocol, the fluctuations in the position of the weight will increase.}. This differs from the results of \cite{HorOpp11}, who show that such transitions are irreversible when considering  (almost) deterministic work extraction, rather than average work.

\subsection{Proof of optimality}  \label{s:optimality}
We now show that the above protocol is optimal. In particular, we will argue that the maximum free energy that can be extracted from a transition between diagonal states  $\rho$ and $\sigma$ is $F_\beta(\rho) - F_\beta(\sigma)$, using a modification of our proof of the second law (in Section \ref{s:second_law}).

Suppose that we have a protocol which extracts average work $F_\beta(\rho) - F_\beta(\sigma) + \delta$ (where $\delta > 0$) when the system is transformed from $\rho$ and $\sigma$. We can then use the above protocol to return the state from $\sigma$ to $\rho$ in a finite number of steps, extracting work  $F_\beta(\sigma) - F_\beta(\rho) - \epsilon$, where we choose the number of thermal qubits  such that $ \epsilon$  is in the range $0 < \epsilon < \delta/2$. The net effect is that a positive average work $\geq \delta/2$ is extracted into the weight, and the system begins and ends the combined procedure in the same state $\rho$.

Considering total energy conservation for the combined protocol, we note that the average energy of the bath must have decreased by at least $\delta/2$, which means that the bath's entropy must also have decreased by a finite amount. However, the system does not change in entropy, and the  change in entropy of the weight can be made as small as desired by choosing its initial state to be a sufficiently broad pure state (see Appendix \ref{a:work_independence}). Hence entropy conservation cannot be satisfied by our hypothetical protocol, and thus no protocol exists which extracts more average work than the free energy change.

\section{Extracting work from coherence}\label{s:coherence}
So far all of our discussions have involved only states which are diagonal in the energy eigenbasis and the protocol presented involved shifting probability only between these energy eigenstates. We can of course consider more general states, and ask the question of how to extract the maximum amount of work from such states.

\subsection{Motivating example}
As a motivating example, let us consider a system with Hamiltonian $H_S = E_S\proj{1}{S}$ prepared initially in the pure state
\begin{equation}\label{e:pure}
	\kets{\psi}{S} = \sqrt{1-p_{eq}}\kets{0}{S} + \sqrt{p_{eq}}\kets{1}{S}
\end{equation}
This state has the same average energy as the thermal state $\tau_\beta(H_S)$, however, since this state is pure, it has zero entropy and hence its free energy differs from that of the thermal state. Thus one would like to extract work from this state, with the limit on extractable work given by
\begin{equation} \label{e:opt work}
	W \leq F(\psi) - F(\tau_\beta(H_S)) = TS_c(p_{eq})
\end{equation}
Previously we were able to extract work from single copies of diagonal states. However,  we show in the next section that it is impossible to  extract work from a single copy of $\kets{\psi}{S}$ given the assumptions of the framework. This is similar to the analysis of \cite{HorOpp11} where they also found that the state \eqref{e:pure} is useless for (almost) deterministic work extraction.

However, it is only a single copy which is a useless resource; two copies can be used to extract work, although the amount is  less than that given by (\ref{e:opt work}). The important point is the way in which work is extracted, which will suggest a different way in which one can understand an asymptotic protocol (in the number of system qubits) able to extract the optimal amount of work.

Consider two copies of the state $\kets{\psi}{S}$, which written out in full is given by
\begin{multline}
	\kets{\psi}{S}^{\otimes 2} = (1-p_{eq})\kets{0}{S}\kets{0}{S} + \sqrt{p_{eq}(1-p_{eq})}\Big(\kets{0}{S}\kets{1}{S}\\+\kets{1}{S}\kets{0}{S}\Big) + p_{eq}\kets{1}{S}\kets{1}{S}
\end{multline}
Let us consider that we now dephase this state in the energy eigenbasis, so called collective-dephasing. We can achieve this in two ways, either by allowing the state to evolve under its Hamiltonian freely for some unknown duration of time, or more systematically by applying controlled-$Z$ operations between maximally mixed (therefore degenerate) bath qubits and the system. Note that both of these are energy conserving unitaries. In either case, at the end of the dephasing the state $\rho$ is
\begin{multline}
	\rho = (1-p_{eq})^2 \kets{0}{S}\kets{0}{S} \bras{0}{S}\bras{0}{S} + 2p_{eq}(1-p_{eq})\kets{\psi^+}{S}\bras{\psi^+}{S} \\+ p_{eq}^2\kets{1}{S}\kets{1}{S}\bras{1}{S}\bras{1}{S}
\end{multline}
where $\kets{\psi^+}{S} = \tfrac{1}{\sqrt{2}}\left(\kets{0}{S}\kets{1}{S}+\kets{1}{S}\kets{0}{S}\right)$. We note that since $\kets{0}{S}\kets{1}{S}$ and $\kets{1}{S}\kets{0}{S}$ have the same energy, the definite phase between them is preserved. The important point is that when considering the combined system we find that we have a two-dimensional degenerate subspace and within this subspace we are in the pure state $\kets{\psi^+}{S}$, and \emph{not} the orthogonal state $\kets{\psi^-}{S} \equiv \tfrac{1}{\sqrt{2}}\left(\kets{0}{S}\kets{1}{S}-\kets{1}{S}\kets{0}{S}\right)$. Thus we have knowledge that we are with certainty in one of two orthogonal states and we can use this information to extract work from a thermal bath by allowing it to `isothermally expand' into the full subspace. Randomising these two states brings the pair of qubits to their thermal state. We can thus apply the analogue of the protocol in Sec.~\ref{ss:maxwell demon}, coupling only the two states inside the degenerate subspace to the thermal bath, and it follows immediately that the final average energy of the weight, in the asymptotic limit, is
\begin{equation}
	\Exp{E}_w = 2p_{eq}(1-p_{eq})T\log 2
\end{equation}
Except for the limiting case $p_{eq} = 0$, corresponding to the unphysical condition $E_S \to \infty$, this value is strictly less than the change in free energy of two system qubits,
\begin{equation}
	2p_{eq}(1-p_{eq})T\log 2 < 2TS_c(p_{eq})
\end{equation}
This inequality is due to the fact that the collective-dephasing changed the entropy of the system but extracted no work in the process. One may wondering therefore why this operation is considered in the first place. In Appendix \ref{a:coherence} we show that we can apply the same analysis, including the collective-dephasing, to $n$ copies of $\kets{\psi}{S}$ and extract, in the asymptotic limit $n \to \infty$, work equal to the free energy change $n T S_c(p_{eq})$. This is possible since, as we also show, collective-dephasing increases the entropy of $n$ systems by an amount which grows as $\log(n+1)$, which becomes negligible in the asymptotic limit. We thus see that coherence between states of different energy in fact plays no thermodynamic role asymptotically in work extraction\footnote{ In the other direction, that of creating non-thermal states from a supply of work, in \cite{BraHorOpp11} it was shown that a phase reference (e.g. a large coherent state) is consumed at a sub-linear rate in this process.}.

Altogether this highlights an essential difference between diagonal and non-diagonal states;  When
$E_S \neq 0$,  the state of multiple qubits in a diagonal state will be maximally mixed inside each energy eigenspace. However, for non-diagonal states, these subspaces take on non-trivial states, and work can be extracted from them by allowing them to `isothermally expand' into the full subspace.

\subsection{General work extraction protocol}
We now present a general strategy to extract work from $n$ copies of an arbitrary state $\rho$, of arbitrary dimension $d$. It is helpful to divide the protocol into three steps:
\begin{enumerate}
\item Collectively dephase the state $\rho^{\otimes n}$ in the joint energy eigenbasis. \label{i:dephase}
\item Allow the resulting state to `isothermally expand' inside each energy eigenspace, until it equals $\omega^{\otimes n}$, where $\omega$ is the state obtained by dephasing $\rho$ in its energy eigenbasis. \label{i:expand}
\item Individually transform each copy of $\omega$ into $\tau_\beta(H_S)$. \label{i:transform}
 \end{enumerate}
It follows from the results of Sec.~\ref{ss:qudits} that steps (\ref{i:expand}) and (\ref{i:transform}) can be performed whilst extracting an average amount of work equal to the free energy change. The free energy change in the first step is not captured as work, but is simply lost. However, we show in the Appendix \ref{a:coherence} that the amount of work lost in this step is at most $T (d-1) \log(n+1)$.

The average amount of work extracted per copy of the original system is therefore
\begin{equation}
\delW \geq  F_\beta(\rho) - F_\beta(\tau_\beta(H_S)) - \frac{T (d-1)}{n} \log(n+1).
\end{equation}
In the limit as $n\rightarrow \infty$, we therefore extract an average work per copy of $\rho$ equal to its free energy change, as desired.

If we only have a single copy of the system, the amount of work we can extract from $\rho$ is  given by the free energy change of the decohered state, $F_\beta(\omega) - F_\beta(\tau_\beta(H_S))$. This can be obtained by first dephasing $\rho$ into $\omega$ and then using the protocol in Sec.~\ref{ss:qudits}.  We now show that this is optimal, following the approach given in \cite{HorOpp11}.

 Firstly, note that as we require the average work extracted by a protocol to be independent of the initial state of the weight, we are free to choose that state however we like -- here we choose it to be a very narrow wavepacket centred on zero. Now consider a decomposition of the total state space into subspaces, each of which has total energy (of the system, bath and weight) close to one of the energy eigenvalues of the system and bath\footnote{In particular the $i^{\textrm{th}}$ subspace  corresponds to the total energy $E$ lying in the range $\frac{E_i-E_{i-1}}{2} \leq E \leq \frac{E_{i+1}-E_{i}}{2}$, where $E_i$ are the energy eigenvalues of the system and bath. Furthermore we choose the width of the weight's initial wavepacket to be narrower than the smallest subspace.}. Note that any work extraction protocol can be followed by a transformation which decoheres the state with respect to these total energy subspaces, without affecting the average work extracted. However, this decohering operation commutes with the unitaries used in the protocol, so we can move it to the beginning of the protocol without changing the work extracted. At the beginning of the protocol, this operation has the sole effect of decohering the system in its energy eigenbasis (changing $\rho$ to $\omega$). Hence the protocol extracts the same amount of work as it would have if it had operated on $\omega$.

\section{Perspective using `virtual temperatures'}\label{s:virtual}
Consider the model of the smallest possible heat engines \cite{YouMahOba09,BruLinPop11}, consisting of two qubits, one with energy spacing $E_1$ in thermal contact with a bath at temperature $T_1$, and a second with energy spacing $E_2$ in contact with a bath at $T_2$. As a composite system this pair of qubits is not in a global thermal state. Nevertheless, let us assign a `virtual temperature' to each transition via the identification \cite{BruLinPop11}
\begin{equation}
	\tfrac{p(e)}{p(g)} = e^{-\beta_v (E_e-E_g)}
\end{equation}
where $p(e)$ and $p(g)$ are the probability of being in the `excited' state (the state of higher energy) and ground state (the state of lower energy) respectively, with $E_e$ and $E_g$ being their energy eigenvalues, and $\beta_v = 1/T_v$ is the `inverse virtual temperature'. For the case in hand, the virtual inverse temperature for the states $\kets{0}{1}\kets{1}{2}$ and $\kets{1}{1}\kets{0}{S}$ is
\begin{equation}
	\beta_v = \frac{S_c'(p_1)-S_c'(p_2)}{E_1-E_2} = \frac{E_1 \beta_1 - E_2 \beta_2}{E_1-E_2}
\end{equation}
where we have denoted the probability for qubit $i$ to be in the excited state by $p_i$, and we have used the relation $S_c'(p) = \beta E$ for a thermal state.

The motivation for introducing the concept of virtual temperatures is that for the smallest machines it gives a concise characterisation of their behaviour; the function of the machine is determined by the virtual temperature -- refrigerator, heat pump or heat engine, if colder, hotter, or negative temperature respectively; Carnot efficiency is approached when the virtual temperature approaches the external temperature for the refrigerator and pump, and when approaching minus infinite temperature for the heat engine.

It is the final machine mentioned, the heat engine, which is relevant here, since this is the machine which produces work. We can think of the system qubit and bath qubit together to be a two-qubit heat engine and that the operating point of the protocol is the Carnot point of the engine.

Let us return to the motivating example of Section \ref{s:motivating example}. Our system qubit had spacing $E_S$ with excited state probability $p$, whilst the bath qubit had spacing $E_B$ with excited state probability  $q$. Let us rename the state $\kets{1}{B}\kets{0}{S} = \kets{0}{v}$ and $\kets{0}{B}\kets{1}{S} = \kets{1}{v}$. With this renaming, the interaction \eqref{e:U mot} becomes
\begin{equation}
	\kets{1}{v}\kets{x}{w} \leftrightarrow \kets{0}{v}\kets{x+E_v}{w}
\end{equation}
where $E_v = E_S - E_B$ is the spacing of the virtual qubit. Thus from this perspective it is explicit that the unitary is the one which exchanges energy from the virtual qubit into the weight, analogous to the Hamiltonian of the smallest heat engine \cite{BruLinPop11}. The operating point of the protocol was to choose $E_B$ such that $q = p-\delp$. In this regime, to first order, the inverse virtual temperature becomes
\begin{equation}
	\beta_v \simeq \delp\frac{ S_c''(p)}{E_S-TS_c'(p)}
\end{equation}
where $S_c''(p) = -\tfrac{1}{1-p}-\tfrac{1}{p} \leq 0$. Thus the inverse virtual temperature becomes vanishingly small and negative as $\delp \to 0$, which is exactly the Carnot limit of the small heat engine.

The picture that one obtains from this perspective is that the reason why any state which is not a thermal state at temperature $T$ is a resource is because the \emph{composite system of resource state + thermal state} contain non-trivial virtual temperatures. One can extract work from a system approaching its free energy when the only exchanges between the composite system and weight involve virtual qubits whose inverse temperatures become negatively infinite.

Finally, we conclude by noting the protocol for systems with many energy levels becomes easier to explain using this perspective. A non-thermal state will have transitions (i.e. pairs of levels) at many different temperatures, and each pair needs to be bought to the same temperature in the thermalisation process. In the protocol the bath qubits interact only with a single pair of levels at a time. To approach reversibility it must be the case that any systems which interact only differ in temperature by an infinitesimal amount, but this is exactly the condition enforced by choosing the renormalised probability of the pair to be close to the bath qubit populations. Finally, the unitary which is applied is the basic thermalising unitary and the energy exchanged in the limit is precisely the free energy.

\section{Conclusions}\label{s:conclusions}
To summarise, in this paper we have analysed the fundamental laws of thermodynamics using the recently proposed concept of resource theories. We proposed an alternative resource theory which differs in certain aspects from the original one. In this conceptual framework we have proved a number of results. First we proved the second law of thermodynamics. Second we proved that for systems that are diagonal in the energy eigenbasis, but not at equilibrium with the thermal bath, we can extract work equal to the free energy by processing each system \emph{individually}, hence showing the free energy is a relevant quantity for individual systems. Third, we proved that when systems are not diagonal in the energy eigenbasis, i.e. they have coherences between different energy eigenstates, it is not possible to extract work equal to the free energy by processing them individually. Work equal to the free energy can however be extracted if we allow for collective processing of such systems (in the asymptotic limit of many copies). 

A number of important questions open immediately. First of all, whether the assumptions which underlie our resource theory framework are too restrictive. In particular, this may be the reason why we cannot extract the entire free energy by processing single systems with coherences between different energy levels. Indeed, given our assumptions we are very limited in how we can access and manipulate the relative phase between two different energy levels, hence we cannot interfere them in a controlled way. As such, for all practical purposes the different energy levels behave as if they were decohered; it is precisely the free energy of this decohered state that is the maximum we can extract in our framework. To be able to interfere different energy levels, i.e. to control their relative phase, we need a ``frame of reference" for it. One way to look at this is that we need some other resource with coherent superpositions between energy levels \footnote{Note that the weight cannot be used for this purpose because of our assumption that the work extracted must be independent of its initial state}. Another way to express this is that we essentially need a clock: the phase between two energy levels varies in time, so a ``frame of reference" for it is essentially a clock. When collectively processing many copies the different systems act effectively as reference systems for each other. For individual processing however we essentially need to add an external clock.  In our present framework we did not include such a clock since we were not able to guarantee that one cannot unwittingly cheat by using the clock itself as a source of free energy. It is conceivable however that a more careful formulation of the resource theory would permit integrating a clock in a consistent and controlled manner. In this case, it may be possible that work equal to the free energy can be extracted from all quantum states by only processing individual copies. 

Second, while in the above we asked if we can relax some of the constraints imposed in the resource theory, for example by including a clock, we must also enquire whether we may indeed need to strengthen some of the constraints. In particular, one may be concerned that we have used unitary transformations as opposed to Hamiltonian evolution. Unitaries imply controlled evolution, i.e. turning on and off interactions at precise time intervals. One may wonder whether this does not require some free energy to achieve. If this is the case, then it is only an illusion that we have extracted work equal to the free energy from the system, since we had to pay to implement the process. Although such a detailed analysis is missing, the fact that all of the results obtained within the resource theory framework so far are consistent with our expectations is an implicit indication that the assumptions are indeed justified.

To conclude, the resource theory framework seems to be a natural and powerful way to approach thermodynamics. It has already delivered significant results and we believe that further investigation along these lines will lead to a much deeper understanding of the foundations of thermodynamics. 

\begin{acknowledgments}  SP and PS acknowledge support from the European project (Integrated Project ``Q-ESSENCE"). SP acknowledges support from the ERC (Advanced Grant ``NLST''), and the Templeton Foundation. AJS acknowledges support from the Royal Society. PS is grateful to Lidia del Rio, Johan {\AA}berg, Joe Renes, Philippe Faist, Renato Renner, Oscar Dahlsten and Luis Masannes for interesting discussions.
\end{acknowledgments}

\providecommand{\bysame}{\leavevmode\hbox to3em{\hrulefill}\thinspace}
\providecommand{\MR}{\relax\ifhmode\unskip\space\fi MR }
% \MRhref is called by the amsart/book/proc definition of \MR.
\providecommand{\MRhref}[2]{%
  \href{http://www.ams.org/mathscinet-getitem?mr=#1}{#2}
}
\providecommand{\href}[2]{#2}

\appendix
\section*{APPENDICES}
\section{Independence of the work on the initial state of the weight} \label{a:work_independence}

In each step of our protocols in which work is extracted, the state of a thermal qubit is swapped with a pair of energy levels in the system, with a corresponding translation of the weight in order to conserve total energy. Consequently, each work-extraction step can be represented by a unitary transformation of the form
\begin{equation}
U = \sum_{ij} \Pi_{ij} \ket{i}\bra{j} \otimes{\Gamma_{E_j - E_i}}
\end{equation}
where the states $\{ \ket{i} \}$ are an orthonormal basis of energy eigenstates for the system and the relevant portion of the bath, such that $(H_B + H_S) \ket{i}= E_i  \ket{i}$, $\Gamma_a$ is the translation operator on the weight, and $\Pi_{ij}$ is a permutation matrix. Furthermore, apart from the initial decohering of non-classical states (discussed in Section \ref{s:coherence}), which is entirely independent from the weight, the entire protocol can be cast in this form.

It is easy to see that all such unitaries commute with translations on the weight. We now show that the average work extracted by a unitary of this form does not depend on the initial state of the weight (even if it is initially correlated with the state of the system).  Let us denote the initial state of the system, bath and weight by the density operator $\rho$. The average work extracted is given by
\begin{eqnarray}
\Exp{W} &=& \tr \left( H_W  U \rho U^{\dag} \right) -  \tr \left( H_W \rho \right), \nonumber \\
&=& \tr \left(  \left(U^{\dag} H_W  U - H_W \right) \rho \right) \label{eq:workindependence}
\end{eqnarray}
where $H_W = \openone_{SB} \otimes \hat{x}_w $ is the Hamiltonian of the weight (defined for convenience as an operator on the entire system). Now note that
\begin{eqnarray}
U^{\dag}  H_W U  &=& \sum_{ijkl}  \Pi_{kl}\Pi_{ij} \ket{l}\braket{k}{i}\bra{j} \otimes \left(\Gamma_{E_k - E_l} \hat{x} \Gamma_{E_j - E_i} \right) \nonumber \\
&=& \sum_{ijl}  \Pi_{il}\Pi_{ij} \ket{l}\bra{j} \otimes \left(\Gamma_{E_i - E_l} \hat{x} \Gamma_{E_j - E_i} \right) \nonumber \\
&=& \sum_{ij}  \Pi_{ij} \ket{j}\bra{j} \otimes \left( \Gamma_{E_i - E_j} \hat{x} \Gamma_{E_j - E_i} \right)  \nonumber \\
&=& \sum_{ij}  \Pi_{ij} \ket{j}\bra{j} \otimes \left(\hat{x} + \left(E_j - E_i\right) \openone_w \right) \nonumber \\
&=& H_W + \sum_{ij}  \Pi_{ij} \left(E_j - E_i\right) \ket{j}\bra{j} \otimes \openone_w
\end{eqnarray}
where in the third line we have used the fact that for permutation matrices $\Pi_{il} \Pi_{ij}= \Pi_{ij} \delta_{jl}$  (where $\delta_{jl}$ is the Kronecker delta function), and in the fourth line we have used the fact that $\Gamma_{-a} \hat{x} \Gamma_{a} = \hat{x} + a \openone_w$. Inserting this expression in
(\ref{eq:workindependence}), we obtain
\begin{equation}
\Exp{W} = \tr_{SB} \left( \Big(\sum_{ij}  \Pi_{ij} \left(E_j - E_i\right) \ket{j}\bra{j}\Big)  \rho_{SB} \right)
\end{equation}
where $\rho_{SB} = \tr_W (\rho)$ is the reduced density matrix of the system and bath. Hence the average amount of work extracted is independent of the initial state of the weight as desired.

Note that the most general energy-conserving unitary transformations which commute with all translations on the weight have the form
\begin{equation} \label{e:general_unitary}
U = \sum_{ij} u_{ij} \ket{i}\bra{j} \otimes{\Gamma_{E_j - E_i}}
\end{equation}
where $u_{ij}$ is an arbitrary unitary matrix. However, for such unitaries the work extracted would generally depend on the initial state of the weight.

\section{Proof of second law of thermodynamics}  \label{a:second_law}
In this appendix, we provide further details for our proof of the second law in Section \ref{a:second_law}.

We first argue that any reduction in the average energy of an initially thermal state (with positive temperature) must also yield a reduction in its entropy. We will only need to use the fact that a thermal state is the maximal entropy state with a given average energy. If a system starts in a thermal state and is transformed to a final state with fixed average energy, the entropy change $\Delta S_B$ will be maximised when the final state is also thermal. In the case where the average energy decreases, and the initial state has positive temperature, the final state will be a thermal state of lower temperature, and thus lower entropy.

We now show that the entropy change of the weight in any particular protocol may be made as small as desired by choosing its initial state to be a sufficiently broad pure state (e.g. a state $\ket{\psi_L}$ with wavefunction $\psi_L(x)$ equal to $1/\sqrt{2L}$ for $|x|\leq L$, and 0 otherwise, where $L \gg 1$). First note that the requirement that transformations are translationally invariant on the weight means that they have the form given in  (\ref{e:general_unitary}). When applied to a thermal state of the bath, the effect of the protocol on the weight is to transform it into a finite mixture of translations of the initial state $\rho_w = \sum_{ij} p_{ij} \Gamma_{E_{i} -E_{j}} \proj{\psi_L}{w} \Gamma_{E_{j} -E_{i}}$, where $p_{ij}$ are probabilities and $i$ and $j$ label energy eigenstates (of energies $E_i$ and $E_j$) of the bath. If the thermal systems used in the protocol have total dimension $d$, and the initial state of the weight is pure,  then the final state of the weight lies  in a  $d^2$-dimensional subspace spanned by the states $ \Gamma_{E_{i} -E_{j}} \ket{\psi_L}{w}$ for $i,j \in \{ 1,2, \ldots d\}\}$.

As the final state of the weight lives in a finite dimensional subspace, its entropy can be shown to be continuous due to Fannes' inequality \cite{NieChu00}.
\begin{equation}
|S(\rho_w) - S(\proj{\psi_L}{w})| \leq D \log \left( \frac{d^2}{D} \right)
\end{equation}
where $ D$ is the trace distance between $\rho_W$ and $\proj{\psi_L}{w}$. We can make $ D$ as small as we like by taking sufficiently large $L$ and therefore make the entropy change as small as we like. For the case in hand, a straightforward calculation shows that $D \leq \sqrt{\frac{a}{L}}$, where $a = \max_{ij}|E_i-E_j|$ is the maximum energy gap of the system-bath Hamiltonian. 

Note that all of the above arguments apply equally well if we consider a system in addition to the bath which is initially in a mixture of energy eigenstates, as considered in Section \ref{s:optimality}.

\section{Motivating example details}\label{a:motivating example}
In this appendix we shall provide more details into the example of Section \ref{s:motivating example}, where a state undergoes a transformation and in the process supplies an amount of work $W$.

Consider the state
\begin{equation}
	\rho_S = (1-p)\proj{0}{S} + p \proj{1}{S}
\end{equation}
with Hamiltonian $H_S = E_S \proj{1}{S}$, along with a thermal qubit
\begin{equation}
	\tau_\beta(H_B) = (1-r)\proj{0}{B} + r\proj{1}{B}
\end{equation}
with Hamiltonian $H_B = E_B\kets{1}{B}$, where $E_B = T S_c'(r)$. Let us bring in a weight in the state $\kets{0}{w}$, and apply the unitary transformation $U$, which interchanges
\begin{equation}
	\kets{0}{B}\kets{1}{S}\kets{x}{w} \leftrightarrow \kets{1}{B}\kets{0}{S}\kets{x+E_S-E_B}{w}
\end{equation}
whilst leaving all other orthogonal states unchanged. Applying this unitary to the combined system we find that
\begin{multline}\label{e:U example}
	\sigma_{BSw} = U\tau_\beta(H_B)\otimes \rho_S \otimes \proj{0}{w}U^\dagger \\ =  rp\kets{1}{B}\kets{1}{S}\kets{0}{w}\bras{1}{B}\bras{1}{S}\bras{0}{w} \\ + r(1-p)\kets{0}{B}\kets{1}{S}\kets{E_B-E_S}{w}\bras{0}{B}\bras{1}{S}\bras{E_B-E_S}{w} \\
	+ (1-r)p\kets{1}{B}\kets{0}{S}\kets{E_S-E_B}{w}\bras{1}{B}\bras{0}{S}\bras{E_S-E_B}{w} \\
	+(1-r)(1-p)\kets{0}{B}\kets{0}{S}\kets{0}{w}\bras{0}{B}\bras{0}{S}\bras{0}{w}
\end{multline}
From this one can finds that the final reduced state of the system and bath qubit are $\rho' = \tr_{Bw} \sigma =\tau_\beta(H_B)$ and $\tau'~=~\tr_{Sw}\sigma = \rho_S$ respectively. Thus we see that at the level of reduced states these two systems have swapped states. The change in free energy of the system $\Delta F_\beta$ is
\begin{align}
	\Delta F_\beta &= F_\beta(\rho) - F_\beta(\tau_\beta(H_B)) \nonumber \\
	&= (p E_S - T S_c(p)) - (r E_S - T S_c(r)) \nonumber \\
	&= (p-r)E_S -T(S_c(p)-S_c(r))
\end{align}

Moving on to the weight its final reduced state $\Omega_w'$ is
\begin{multline}\label{e:Lambda L}
	\Omega_w' = (rp+(1-r)(1-p))\proj{0}{w}\\+ r(1-p)\proj{E_B-E_S}{w}\\  + (1-r)p\proj{E_S-E_B}{w}
\end{multline} 
The change in average energy of the weight $\Delta \Exp{E}_w$ is thus
\begin{align}
	\Delta \Exp{E}_w &= \tr(H_w \Omega_w) - \tr(H_w \proj{0}{w}) \nonumber \\
	&= (p-r)(E_S - E_B) \nonumber \\
	&= (p-r)(E_S - TS_c(r))
\end{align}
Now, let us take $r=p-\delp$ and expand to second order in $\delp$ both $\Delta F_\beta$ and $\Delta \Exp{E}_w$,
\begin{align}
	\Delta F_\beta &= \delp(E_S - T S_c'(p)) + \delp^2 \tfrac{T S_c''(p)}{2} + \mathcal{O}(\delp^3) \\
	\Delta \Exp{E}_w &= \delp(E_S - T S_c'(p)) + \delp^2 T S_c''(p) + \mathcal{O}(\delp^3)
\end{align}
The change in free energy and the change in average energy of the weight thus coincide to first order in $\delp$ and differ only by a factor of 2 in the second order term. Since $S_c''(p) < 0$, we see immediately that $\Delta \Exp{E}_w \leq \Delta F_\beta$, so that the change in free energy is always larger than the change in average energy of the weight, as we would expect.

\section{Calculating the average energy and spread}\label{a:weight}
In this appendix we will show how to arrive at equation \eqref{e:asymptotic ladder} for the asymptotic state of the ladder. To do so let us consider first those instances where the system qubits starts in state $\kets{0}{S}$. Given this we shall then calculate the expected energy of the weight and its variance in the limit of large $N$ to first order, from which we shall be able to conclude that it approaches a definitive value in the asymptotic limit at the desired energy. A analogous method can also be employed assuming the system starts instead in the state $\kets{1}{S}$, with the desired result also obtained.

By calculating the first few stages of the protocol and then by induction it is easy to show that the expected energy of the weight at the end of the protocol is
\begin{equation}\label{e:Ew0}
	\Exp{E}_w^{[0]} = -p_{eq} E_S + p E_B^{(1)} + \sum_{k=1}^N (q^{(k)} - q^{(k-1)})E_B^{(k)}
\end{equation}
where the superscript $[0]$ is a reminder that this is the case for the system starting in the state $\kets{0}{S}$, and as defined in the main text we have
\begin{align}
	q^{(k)}& = p + \tfrac{k}{N}(p_{eq}-p)& E_B^{(k)} &= \tfrac{1}{\beta}\log\left(\tfrac{1-q^{(k)}}{q^{(k)}}\right)
\end{align}
By expanding the logarithms in powers of $\tfrac{k}{N}$ and denoting $a = (p_{eq}-p)$, after some basic manipulations this can be re-written as
\begin{multline}
	\Exp{E}_w^{[0]} = -p_{eq} E_S + p E_B^{(1)} + \tfrac{a}{\beta}\log\left(\tfrac{1-p}{p}\right) \\
	+ \frac{a}{\beta N}\sum_{k=1}^N \sum_{n=1}^{\infty} \left(-\frac{1}{n}\left(\frac{ak}{(1-p)N}\right)^n +\frac{1}{n}\left(\frac{-ak}{pN}\right)^n \right)
\end{multline}
The advantage of performing this expansion is that we can now interchange the order of summation and note that the only term which depends upon $k$ is the term $k^n$ which is common to both terms in the sum. We can thus use the expansion
\begin{equation}
	\sum_{k=1}^N k^n= \frac{N^{n+1}}{n+1} + \frac{N^n}{2} + \mathcal{O}\left(N^{n-1}\right)
\end{equation}
to write
\begin{multline}
		\Exp{E}_w^{[0]} = -p_{eq} E_S + p E_B^{(1)} + \tfrac{a}{\beta}\log\left(\tfrac{1-p}{p}\right) \\
		+ \frac{1}{\beta} \sum_{n=1}^{\infty} \left(\frac{a^{n+1}}{n}\left(\frac{1}{(-p)^n}-\frac{1}{(1-p)^n} \right)\left(\frac{1}{n+1} + \frac{1}{2N}\right)\right) \\+ \mathcal{O}\left(\tfrac{1}{N^2}\right)
\end{multline}
This sum can now be evaluated explicitly, and after re-expressing $E_S$ in terms of $S_c'(p_{eq})$, expanding $E_B^{(1)}$ to first order in powers of $\tfrac{1}{N}$, and an amount of manipulation we arrive at
\begin{multline}
	\Exp{E}_w^{[0]} = \tfrac{1}{\beta}\log\left(\tfrac{1-p}{1-p_{eq}}\right) + \tfrac{1}{\beta}(p_{eq}-p)\Big(\tfrac{1}{2}\big(S_c'(p_{eq})\\-S_c'(p)\big) -\tfrac{1}{1-p}\Big)\tfrac{1}{N} + \mathcal{O}\left(\tfrac{1}{N^2}\right)
\end{multline}
thus demonstrating that the weight is peaked around the point expected when the system starts off in the state $\kets{0}{S}$ with an offset which vanishes in the asymptotic limit.

Considering the case now when the system starts off in the state $\kets{1}{S}$ and again calculating the first few stages of the protocol and then by induction it can be shown that
\begin{equation}
	\Exp{E}_w^{[1]} = \Exp{E}_w^{[0]} + E_S - E_B^{(1)}
\end{equation}
Thus by re-expressing $E_S$ in terms of $S_c'(p_{eq})$ and expanding $E_B^{(1)}$ to first order it is straightforward given the previous results to show that
\begin{multline}
	\Exp{E}_w^{[1]} = \tfrac{1}{\beta}\log\left(\tfrac{p}{p_{eq}}\right) + \tfrac{1}{\beta}(p_{eq}-p)\Big(\tfrac{1}{2}\big(S_c'(p_{eq})\\-S_c'(p)\big) +\tfrac{1}{p}\Big)\tfrac{1}{N} + \mathcal{O}\left(\tfrac{1}{N^2}\right)
\end{multline}
which again gives the desired result. Let us move on now to a calculation of the variance $\Delta E_w^{2[i]} = \Exp{E^2}_w^{[i]} - \Exp{E}_w^{2[i]}$ of each of the two packets, where for simplicity we will neglect the finite width of the initial wavepacket. Again by induction it can be shown that the spread of both packets is identical (and hence we shall drop the superscript $[i]$) and equal to
\begin{equation}
	\Delta E_w^2 = \sum_{k=1}^N q^{(k)}(1-q^{(k)})(E_B^{(k+1)}-E_B^{(k)})
\end{equation}
Let us consider the $k^{th}$ term in this sum, multiplied by $N$. Defining $x = \tfrac{k}{N}$ and expanding this summand in terms of $\tfrac{1}{N}$ we find
\begin{multline}
	q^{(k)}(1-q^{(k)})(E_B^{(k+1)}-E_B^{(k)}) \\= \frac{T^2(p_{eq}-p)^2}{(p+x(p_{eq}-p))(1-p-x(p_{eq}-p))N^2} + \mathcal{O}\left(\tfrac{1}{N^3}\right)
\end{multline}
Thus for large $N$ the leading order contribution to this sum becomes equal to
\begin{equation}
	\Delta E_w^2 \simeq \frac{1}{N}\int_0^1  \frac{T^2(p_{eq}-p)^2 dx}{(p+x(p_{eq}-p))(1-p-x(p_{eq}-p))}
\end{equation}
Evaluating this integral explicitly we find finally that
\begin{equation}
	\Delta E_w^2 = \frac{T^2 (p_{eq}-p)(S_c'(p_{eq}-S_c'(p))}{N} + \mathcal{O}\left(\frac{1}{N^2}\right)
\end{equation}
Thus we see that each peak has a spread of the order of $1/\sqrt{N}$ which vanishes in the asymptotic limit $N\to\infty$. Therefore in this limit we can conclude that the distribution of the ladder approaches a two-peaked distribution, as given by equation \eqref{e:asymptotic ladder}.
\section{Extracting work from diagonal quantum systems of arbitrary dimension}\label{a:qudits}
In this appendix we wish to show that the average work extracted when moving the occupation probability between energy levels is still equal to the free energy in a $d$-dimensional system. It will suffice to consider a qutrit, and moving probability  $\delp$ from the ground state $\kets{0}{S}$ to the second excited state $\kets{2}{S}$.

Let us take our system to be initially in the state
\begin{equation}
	\rho_S = (1-p-q)\proj{0}{S}+q\proj{1}{S} + p\proj{2}{S},
\end{equation}
and thus to end in the state
\begin{equation}
	\rho'_S = (1-p-q-\delp)\proj{0}{S}+q\proj{1}{S}+(p+\delp)\proj{2}{S}.
\end{equation}
where the system Hamiltonian is
\begin{equation}
	H_S = E_1\proj{1}{S} + E_2\proj{2}{S}.
\end{equation}
To achieve this, we append a bath qubit with energy level spacing $E_B$ and ground state population $r$ and apply the unitary transformation which interchanges only the states
\begin{equation}
	\kets{1}{B}\kets{0}{S}\kets{x}{w} \leftrightarrow \kets{0}{B}\kets{2}{S}\kets{x+E_B - E_2}{w}
\end{equation}
This induces on the system and weight the transformation
\begin{multline}\label{e:trit 1}
\rho_S \otimes \proj{0}{w} \xrightarrow{} 					
(1-p-q)(1-r)\kets{0}{S}\kets{0}{w}\bras{0}{S}\bras{0}{w} \\
+q\kets{1}{S}\kets{0}{w}\bras{1}{S}\bras{0}{w} +pr\kets{2}{S}\kets{0}{w}\bras{2}{S}\bras{0}{w} \\
	+p(1-r)\kets{0}{S}\kets{E_{2}-E_B}{w}\bras{0}{S}\bras{E_{2}-E_B}{w} \\
	+(1-p-q)r\kets{2}{S}\kets{E_B-E_{2}}{w}\bras{2}{S}\bras{E_B-E_{2}}{w}
\end{multline}
where we now carry out the final trace over the bath qubit for brevity of presentation. Thus in order that the final reduced state of the system is $\rho'_S$ we find that the ground state population of the bath must be chosen to be
\begin{equation}
	r = \frac{p+\delp}{1-q}
\end{equation}
 The denominator is the total probability to be in either of the states $\kets{0}{S}$ or $\kets{1}{S}$, and hence this expression can be seen as a renormalised version of the qubit case, where the total probability to be in either state always sums to unity. We see that it is the ratio of probabilities which must be close to the ratio of populations of the thermal qubits, which is equivalent to saying that they must be close in temperature (see Sec.~\ref{s:virtual}).

The change in energy of the weight is given by
\begin{eqnarray}
	\delW &=& (p-(1-q)r)(E_2-E_B) \nonumber \\
	&=& -\delp(E_2-E_B)
\end{eqnarray}
thus using the fact that $E_B = \tfrac{1}{\beta}S_c'(r)$ and expanding to first order in $\delp$ we find that the work extracted from the system is given by
\begin{equation}
	\delW \simeq -\delp\left(E_2 -\tfrac{1}{\beta}S_c'\!\!\left(\tfrac{p}{1-q}\right)\right)
\end{equation}
which a straightforward calculation verifies is equal to the change in free energy $F(\rho_S)~-~F(\rho_S')$ to first order in $\delp$. If we do a sequence of $N$ such transitions between the states $\rho$ and $\sigma$, in each of which $\delp = \mathcal{O}(1/N)$, then we will obtain $\delW = F(\rho_S)~-~F(\rho_f) + \mathcal{O}(1/N)$. In the asymptotic limit in which the number of thermal qubits $N$ goes to infinity, we will therefore extract average free energy equal to the free energy change.

\section{Work extraction from coherence}\label{a:coherence}
In this appendix we will present the full details for how work can be extracted from coherence. We shall first consider the special case presented in the main text for $n$ systems, and then move on to the general case.

\subsection{Motivating example}

Consider $n$ copies of $\kets{\psi}{S}$, where
\begin{equation}
	\kets{\psi}{S} = \sqrt{1-p_{eq}}\kets{0}{S} + \sqrt{p_{eq}}\kets{1}{S}
\end{equation}

 There will be $n+1$ energy eigenspaces with energies $k E_S$, $k = 0,\ldots, n$. The dimension of the eigenspace with energy $k E_S$ will be $\binom{n}{k}$. After collective-dephasing in each eigenspace the system will be a pure superposition over all states,  and the probability to be in each eigenspace will be $\binom{n}{k}p_{eq}^k (1-p_{eq})^{(n-k)}$. We then  cause the state to `isothermally expand'  in each degenerate subspace, taking the state from the initial pure state to the final maximally mixed state and in the process extracting $T\log d$ energy from the bath, where $d$ is the dimension of the subspace. In total, the final average energy of the ladder will be
\begin{equation}
	\Exp{E}_w = \sum_{k=0}^n \binom{n}{k}p_{eq}^k (1-p_{eq})^{(n-k)} T\log \binom{n}{k}
\end{equation}
In the asymptotic limit, as $n\to \infty$ it is straightforward to show that
\begin{equation}
	\lim_{n\to\infty}\frac{1}{n}\sum_{k=0}^n \binom{n}{k}p_{eq}^k (1-p_{eq})^{(n-k)}T\log \binom{n}{k} =TH(p_{eq})
\end{equation}
since in the limit the only contribution to the sum comes from $k = np_{eq}$. We thus see that asymptotically the expected amount of work is extracted. There are a couple of comments which are worth mentioning. First we note that due to concentration effects, we know that almost all weight lies in those subspaces with energy $n p_{eq} \pm \mathcal{O}(\sqrt{n})$. We could therefore ignore those subspaces which lie outside this range and obtain, to leading order, the same amount of work. The results obtain however are independent of the concentration effects. Second we recall that the protocol starts with collective-dephasing in the energy eigenbasis. It might appear that this operation, which changes the entropy of the system, throws away work. However, asymptotically we obtain to first order the optimal amount of work, and so we conclude that the amount thrown away must scale sub-linearly in $n$. In fact the amount of entropy thrown away is only $\mathcal{O}(\log n)$, the entropy of the binomial distribution, since the dephasing leaves the system binomially distributed amongst energy eigenspaces which have orthogonal support. This shows that asymptotically the coherence between states of different energy are thermodynamically insignificant.

\subsection{General work extraction protocol}

%adapted this to deal with arbitrary dimensional systems, which is necessary for the general case
Finally, for the case of arbitrary states of arbitrary dimension $d$, we note that the same conclusion can also be drawn. First we need to show that the operation of collective-dephasing never increases the entropy of a collection of systems too much in the asymptotic limit. Let us denote the projectors onto the energy eigenspaces by $\Pi_k$, and the total number of different energies by $K$. After the operation of collective-dephasing the state becomes block diagonal in the energy eigenbasis, namely
\begin{equation}
	\rho^{\otimes n} \mapsto \rho' = \sum_{k=0}^{K-1} \Pi_k \rho^{\otimes n} \Pi_k
\end{equation}
We are interested in the difference in entropy $S(\rho^{\otimes n}) - S(\rho')$, i.e. the entropy generated by this operation. To do so, let us first consider the coherent version of collective-dephasing. Let us introduce an ancillary system $A$ with $K$ states $\kets{z}{A}$ prepared initially in the state $\kets{0}{A}$. Let us also consider the unitary operation $V$
\begin{equation}
	V = \sum_{k,z} \Pi_k \otimes \kets{z+k \mod K}{A}\bra{z}
\end{equation}
Applied to the state $\rho_{SA} = \rho^{\otimes n}\otimes \proj{0}{A}$ this performs the transformation
\begin{equation}
	V \left( \rho^{\otimes n}\otimes \proj{0}{A} \right) V^\dagger = \rho_{SA}' = \sum_{k,k'} \Pi_k \rho^{\otimes n}\Pi_{k'} \otimes \kets{k}{A}\bra{k'}
\end{equation}
It is straightforward to see that $\tr_A \rho'_{SA} = \rho'$ whilst
\begin{equation}
	\tr_S \rho_{SA}' \equiv \rho_A' = \sum_k \tr\left(\Pi_k \rho^{\otimes n}\right) \proj{k}{A} \label{e:rhoA'}
\end{equation}
Considering now the entropies, we have $S(\rho_{SA}')~=~S(\rho^{\otimes n})~=~n S(\rho)$, since the two states differ by the appended pure ancilla and the joint unitary transformation. Thus, from the Araki-Lieb inequality $S(\rho_{SA}) \geq |S(\rho_S) - S(\rho_A)|$ we have
\begin{equation}\label{e:Araki-Lieb}
	n S(\rho) \geq |S(\rho') - S(\rho_A')|
\end{equation}
Next, note that we can rewrite $\rho'$  in the form
\begin{equation}
\rho' = \sum_{k} \tr(\Pi_k \rho^{\otimes n})  \frac{ \Pi_k \rho^{\otimes n}\Pi_{k}}{ \tr(\Pi_k \rho^{\otimes n}) }
\end{equation}
where the sum is restricted to those $k$ for which $\tr(\Pi_k \rho^{\otimes n})~\neq 0$. Comparing this to (\ref{e:rhoA'}), we note that
$S(\rho') \geq S(\rho_A')$, as a mixture of orthogonal mixed states  always has higher entropy than the same mixture of orthogonal pure states. Hence  the absolute value can be omitted on the right-hand-side of equation \eqref{e:Araki-Lieb} and after rearranging we get
\begin{equation}
	 S(\rho_A') \geq S(\rho') - n S(\rho)
\end{equation}
Next, note that $S(\rho'_A) \leq \log K$, since in the worst case $\rho'_A$ is maximally mixed over $K$ states. Hence we find
\begin{equation}
	\log K \geq S(\rho') - n S(\rho)
\end{equation}
The number of different energy levels $K$ is upper bounded by the number of ways of distributing $n$ indistinguishable particles between $d$ distinguishable energy levels, which is given by $\binom{n+d-1}{n}$. Thus
\begin{equation}
 S(\rho') - n S(\rho) \leq \log \binom{n+d-1}{n} \leq  (d-1)\log (n+1)
\end{equation}

This demonstrates that the operation of collective-dephasing on $n$ qubits can only increase the entropy by a logarithmic amount in $n$. The increase \emph{per qubit} is thus of the order $\tfrac{d-1}{n}\log (n+1)$, which vanishes in the asymptotic limit. This result does not follow from any concentration of measure, but simply by the fact that the the number of energy eigenvalues of $n$ non-interacting systems is logarithmic in $n$.

\section{Discrete work storage}\label{a:discrete work}
In this appendix we will show that the energy level spacing of the weight need only become continuous in the asymptotic limit and that for any finite number of bath qubits $N$ it is sufficient to consider the levels to have a minimum spacing $\mathcal{E}$ which scales as $1/N^2$.

Let us begin by recalling how the protocol works. We pick a collection of $N$ bath qubits which have excited state probabilities $q^{(k)}$ given by
\begin{equation}
	q^{(k)} = p + \tfrac{k}{N}(p_{eq}-p)
\end{equation}
and therefore excited state energies $E_B^{(k)} = T S_c'(q^{(k)})$. At each stage of the protocol the weight either gains or loses energy $E_S - E_B^{(k)}$.

Let us now assume instead that the of being a continuous degree of freedom that the weight has a discrete (but infinite) set of states $\kets{k}{w}$ and Hamiltonian
\begin{equation}
	H_w = \sum_{k = -\infty}^{\infty} k \mathcal{E} \proj{k}{w}
\end{equation}
Accordingly, we shall adjust the energy levels of the bath qubits to
\begin{equation}
	\tilde{E}_B^{(k)} = E_B^{(k)} + \varepsilon^{(k)}
\end{equation}
such that
\begin{equation}
	E_S-\tilde{E}_B^{(k)}= m^{(k)} \mathcal{E}
\end{equation}
where the $m^{(k)}$ are integers, and we take the smallest possible $\varepsilon^{(k)}$ possible such that this condition is met, meaning that $|\varepsilon^{(k)}| < \mathcal{E}$.

To proceed, we first note that \eqref{e:Ew0} can alternatively be re-written as
\begin{align}
	\Exp{E}_w^{[0]} &= -p_{eq} E_S + p E_B^{(1)} + \sum_{k=1}^N (q^{(k)} - q^{(k-1)})E_B^{(k)} \nonumber \\
	&= \sum_{k=0}^{N-1} q^{(k)} (E_B^{(k)} - E_B^{(k+1)})
\end{align}
Thus upon making the replacement $E_B^{(k)} \to \tilde{E}_B^{(k)}$ for the case of a discrete weight, the average energy, conditional on the qubit starting in the ground state becomes
\begin{align}
	\Exp{\tilde{E}}_w^{[0]} &= \sum_{k=0}^{N-1} q^{(k)} (\tilde{E}_B^{(k)} - \tilde{E}_B^{(k+1)}) \nonumber \\
	&= \sum_{k=0}^{N-1} q^{(k)} (E_B^{(k)}\! -\! E_B^{(k+1)}) + \sum_{k=0}^{N-1} q^{(k)} (\varepsilon^{(k)}\! -\! \varepsilon^{(k+1)}) \nonumber \\
	&= \Exp{E}_w^{[0]} + \sum_{k=0}^{N-1} q^{(k)} (\varepsilon^{(k)} - \varepsilon^{(k+1)})
\end{align}
It is the final term in the last line which is the error introduced by considering a discrete weight, and thus it is the magnitude of this term which is of interest. This can easily be bounded as follows
\begin{align}
	\left|\sum_{k=0}^{N-1} q^{(k)} (\varepsilon^{(k)} - \varepsilon^{(k+1)})\right|  &\leq \sum_{k=0}^{N-1} q^{(k)} \left|\varepsilon^{(k)} - \varepsilon^{(k+1)}\right| \nonumber \\
	&\leq \sum_{k=0}^{N-1} q^{(k)} 2\mathcal{E}\nonumber \\
	&= \mathcal{E}\left((p+p_{eq})N + (p-p_{eq})\right)
\end{align}
Therefore as long as the product $\mathcal{E}N \to 0$ as $N\to \infty$, then this term will asymptotically vanish. Choosing $\mathcal{E} = \epsilon/N^2$ would ensure that this correction is $\mathcal{O}(1/N)$ and thus the same order as the leading order correction to $\Exp{E}_w^{[0]}$ for finite $N$. An identical analysis holds also for $\Exp{E}_w^{[1]}$, the average energy of the weight conditional on the system starting in the state $\kets{1}{S}$, and therefore we conclude that the results more generally do not rely crucially on the continuous nature of the weight.

\end{document}